\documentclass{PoS}
\usepackage{dsfont}

\title{Topological terms in abelian lattice field
theories\thanks{Based on parallel talks by 
M.~Anosova, C.~Gattringer and D.~G\"oschl.}}

\ShortTitle{Abelian topological terms}

\author{M.~Anosova$^1$, C.Gattringer$^1$, D.~G\"oschl$^1$, T.~Sulejmanpasic$^2$, P.~T\"orek$^1$\\

\vskip2mm

$^1\;$Institute of Physics\footnote{Member of NAWI Graz.} , University of Graz, 8010 Graz, Austria

$^2\;$Department of Mathematical Sciences, Durham University, Durham DH1 3LE, UK

\vskip2mm
        
E-mail: \email{mariia.anosova@uni-graz.at} \\ 
E-mail: \email{christof.gattringer@uni-graz.at} \\
E-mail: \email{daniel.goeschl@uni-graz.at} \\
E-mail: \email{tin.sulejmanpasic@durham.ac.uk} \\
E-mail: \email{pascal.toerek@uni-graz.at} 
}

\abstract{In this contribution we revisit the lattice discretization of the topological charge for abelian lattice field theories. 
The construction departs from an initially non-compact discretization of the gauge fields and after absorbing $2\pi$ shifts of the gauge fields leads to a generalized Villain action that also includes the topological term. The topological charge in two, as well as in four 
dimensions can be expressed in terms of only the integer-valued Villain variables. 
We test various properties of the topological charge and in particular analyze
the index theorem in two dimensions and discuss the Witten effect in 4-d. As an application of our formulation we present
results from a simulation of the 2-d U(1) gauge Higgs model at vacuum angle $\theta = \pi$, where we use a suitable 
worldline/worldsheet  representation to overcome the complex action problem at non-zero $\theta$.}

\FullConference{37th International Symposium on Lattice Field Theory - Lattice2019\\
		16-22 June 2019\\
		Wuhan, China}

\begin{document}

\section{Introduction}

Topological terms are an important and highly interesting aspect of many gauge theories with a wide range of 
applications from low-dimensional effective field theories for condensed matter systems to topological properties
of QCD. However, topological terms also pose a considerable challenge because their physics is non-perturbative in
nature and suitable techniques need to be devised for exploring their role in various systems.
In principle the lattice formulation is a powerful non-perturbative tool but clearly the lattice discretization of topological terms 
is non-trivial because key properties of the various topological terms in the continuum rely on the differentiability
of the gauge fields, a concept obviously absent on the lattice. 

In this contribution we report on a program 
\cite{Gattringer:2018dlw,Goschl:2018uma,Sulejmanpasic:2019ytl,Gattringer:2019yof,WIP1,WIP2}
where we revisit the problem of formulating topological terms for abelian 
gauge theories by exploring the option of using an initially non-compact formulation. This program is based on the lattice discretization which implements the continuum symmetries correctly (see \cite{Sulejmanpasic:2019ytl} for details). Our construction in \cite{Sulejmanpasic:2019ytl} departs from a non-compact formulation of the Abelian gauge theories on the lattice (i.e., the $\mathbb R$-gauge theory), which has a larger center symmetry group than the compact counterparts, namely $\mathbb R$ instead of U(1)\footnote{The center symmetries are not to be confused with the gauge symmetries of the model.}. The center symmetry is then gauged by a discrete 2-form gauge field, whose purpose it is to absorb the shifts from the parameterization of the compact gauge links
that couple to matter fields. The construction
leads to a generalized Villain action for the gauge fields with a natural introduction of the topological charge as a 
functional of the Villain auxiliary variables \cite{Sulejmanpasic:2019ytl,WIP1}.

For a further analysis of the properties of the new definition of the topological charge with our discretization \cite{Sulejmanpasic:2019ytl},
we explored the manifestation of the index theorem on the lattice in \cite{Gattringer:2019yof}. In a short section we here 
report on our results for the index theorem, which indicate that also this conceptually important link  
for the interaction of fermions with topological objects is implemented correctly by our discretization. 
 
In an application of the new formalism for topological terms in 2-d we analyzed the U(1) gauge Higgs system 
with a topological term at vacuum angle $\theta = \pi$ for a 1-component Higgs field 
\cite{Gattringer:2018dlw,Goschl:2018uma}. In the continuum
this model is invariant under charge conjugation for $\theta = \pi$ (and also for the trivial value $\theta = 0$)
and the corresponding $\mathds{Z}_2$ symmetry can be broken spontaneously in 2-d. Our discretization implements this
symmetry exactly on the lattice, a prerequisite for a reliable study of the corresponding phase diagram. 
In addition our formulation allows for a representation in terms of worldlines and worldsheets which overcomes the 
complex action problem, the second necessary ingredient for a Monte Carlo analysis. We briefly summarize the lattice 
discretization as well as the worldline/worldsheet representation, and then discuss recent results of the numerical analysis
of the 2-component model \cite{WIP2} which extends the 1-component analysis 
\cite{Gattringer:2018dlw,Goschl:2018uma}. 

Finally we give a short preview of an upcoming paper \cite{WIP1} where we will discuss in detail the 4-d case
already briefly addressed in \cite{Sulejmanpasic:2019ytl}. The challenge there is to control the monopoles that are 
plaguing the compact formulation. We show that this can be done in a consistent way with our formulation \cite{Sulejmanpasic:2019ytl}
which allows one to construct a suitable topological term in 4-d. We test properties of the topological term and 
show that it correctly reproduces the Witten effect \cite{Sulejmanpasic:2019ytl,WIP1} .  

\section{Discretization of gauge fields and topological terms in 2-d}

We start our presentation of the lattice discretization of U(1) gauge fields with a topological term by defining the 
corresponding 2-d theory in the continuum to set our notation. The continuum action with a topological term is given by 
\begin{equation}
S \; = \; S_g \; - \; i \, \theta \, Q \quad \mbox{with} \quad
S_g \; = \; \frac{1}{2e^2} \int_{\mathds T^2} \!\! d^2x \;  F_{12}(x)^2  \;\; , \;\;\;
Q \; = \; \frac{1}{2\pi} \int_{\mathds T^2} \!\! d^2x \; F_{12}(x)  \; ,
\label{SG_cont}
\end{equation}
where $A_\mu(x) \in \mathds{R}$ and $F_{\mu \nu}(x) \; = \; \partial_\mu\, A_\nu(x) - \partial_\nu \, A_\mu(x)$. The 
integrals are over the 2-torus $\mathds{T}_2$ and differentiable gauge field configurations $A_\mu(x)$ can be classified
by their integer-valued topological charge $Q \in \mathds{Z}$. The gauge fields may be coupled to bosonic or 
fermionic matter fields by replacing the derivatives $\partial_\mu$ 
in the matter field action by the covariant derivative $D_\mu(x) = \partial_\mu + i \, A_\mu(x)$. Under 
gauge transformations with some $\lambda(x) \in \mathds{R}$ matter fields 
$\phi(x)$ transform as $\phi(x) \rightarrow e^{  \, i \, \lambda(x)} \phi(x)$, which 
requires the gauge fields to transform as $A_\mu(x) \rightarrow A_\mu(x) - \partial_\mu \lambda(x)$ in order 
to implement gauge invariance.

Let us now come to our new proposal for the lattice discretization. The first part of this discussion 
is valid for arbitrary dimensions d $\geq 2$ and only when we discuss the topological charge $Q[A]$ 
 will we switch to d = 2 in this section. The topological term in d = 4 dimensions is discussed in a later section below.

To be specific, we work on a d-dimensional lattice of size $V = N_1 \times N _2 \times \, ... \, \times N_{\mathrm{d}}$ where 
d is considered to be the time direction. All lattice fields have periodic boundary conditions, except for 
fermion fields which are anti-periodic in time direction and periodic in all other directions.  
When discretizing the theory on the lattice the derivatives of the matter fields are implemented as nearest neighbor
differences. In order to maintain gauge invariance the products of fields at nearest neighbor sites are 
connected with the link variables $U_{x,\mu}$, giving rise to terms such as 
\begin{equation}
\phi_x^\dagger \, U_{x,\mu} \, \phi_{x + \hat \mu} \quad \mbox{with} \quad U_{x,\mu} \; = \; e^{\, i \, A_{x,\mu}} \; .
\label{nnterms}
\end{equation}
Lattice gauge transformations $\phi_x \rightarrow e^{ \, i \, \lambda_x} \, \phi_x$ and $A_{x,\mu} \rightarrow A_{x,\mu} - 
(\lambda_{x + \hat \mu} - \lambda_x)$ leave the terms (\ref{nnterms}) invariant. In a compact lattice discretization 
the gauge action is constructed directly from the U(1)-valued gauge links $U_{x,\mu}$, giving rise to, e.g., the 
Wilson plaquette action. 

For the discussion of our lattice discretization we decompose the lattice gauge fields 
$A_{x,\mu} \in \mathds{R}$ in the form 
\begin{equation}
A_{x,\mu} \; = \; a_{x,\mu} \; + \; 2\pi \, k_{x,\mu} \quad \mbox{with} \quad a_{x,\mu} \, \in \, [-\pi, \pi] \quad \mbox{and}  
\quad k_{x,\mu} \, \in \, \mathds{Z} \; .
\label{Aparam}
\end{equation}
The values of $a_{x,\mu}$ clearly change the link variables $U_{x,\mu} \; = \; e^{\, i \, A_{x,\mu}} \; = \, e^{\, i \, a_{x,\mu}}$,
while $U_{x,\mu}$ is invariant under the choice of the shifts labelled by $k_{x,\mu} \in \mathds{Z}$. For defining the 
field strength tensor we discretize the continuum expression $\partial_\mu\, A_\nu(x) - \partial_\nu \, A_\mu(x)$ 
in the form of the lattice exterior derivative $(\textrm{d} \, A)_{x,\mu \nu}$ defined as
\begin{equation}
(\textrm{d} \, A)_{x,\mu \nu} \; = \; A_{x + \hat \mu,\nu} \, - \, A_{x,\nu} \, - \, ( A_{x + \hat \nu,\mu} \, - \, A_{x,\mu} ) \; = \; 
(\textrm{d} \, a)_{x,\mu \nu} \; + \; 2 \pi \, (\textrm{d} \, k)_{x,\mu \nu} \; .
\end{equation}
Obviously our candidate for the lattice field strength $(\textrm{d} \, A)_{x,\mu \nu}$ is not invariant under the shifts with 
the variables $k_{x,\mu}$ under which it may pick up additive terms that are integer multiples of $2 \pi$. 
We can restore this invariance by gauging the exterior derivative and defining the field strength for 
our discretization in the form 
\begin{equation}
(\textrm{d}\,A)_{x,\mu \nu} \; + \; 2\pi \, n_{x,\mu \nu} \; = \; 
(\textrm{d}\,a)_{x,\mu \nu} \; + \; 2\pi \, ( \textrm{d}\, k)_{x,\mu \nu} \; + \;  ( \textrm{d}\, n)_{x,\mu \nu} \; ,
\end{equation}
where the plaquette-based fields $n_{x,\mu \nu} \in \mathds{Z}$ are summed over all integers. Clearly 
this prescription eats up the terms $(\textrm{d}\, k)_{x,\mu \nu}$ that emerge from the shifts with the 
$k_{x,\mu}$ that do not change the link variables. We have already remarked that the construction
with removing the contributions from the shifts with a new plaquette-based field $n_{x,\mu \nu}$ 
has the structure of implementing a gauge invariance (not to be confused with the original 
gauge invariance) and for a discussion of various aspects of this gauge invariance 
see \cite{Sulejmanpasic:2019ytl,Gaiotto:2017yup} and references therein.  

Since the contributions of the shifts were removed with the new variables $n_{x,\mu \nu}$ we can drop 
these contributions in our definition of the field strength and finally define it as
\begin{equation}
F_{x,\mu \nu} \; = \; (\textrm{d} \, a)_{x,\mu \nu} \; + \; 2 \pi \, n_{x,\mu \nu} \; .
\label{Fmunudef}
\end{equation}
The gauge action without topological term can thus be defined as 
\begin{equation}
S_g \; = \; \frac{\beta}{2} \sum_{x,\mu < \nu} F_{x,\mu \nu}^{\; 2} \; = \; 
\frac{\beta}{2} \sum_{x,\mu < \nu} \big( (\textrm{d} \, a)_{x,\mu \nu} \; + \; 2 \pi \, n_{x,\mu \nu}\big)^2 \; ,
\label{SG_lat}
\end{equation}
where $\beta = 1/e^2$ and the $n_{x,\mu \nu}$ are summed over all integers. This form of the lattice action is 
known as the {\sl Villain action} \cite{Villain:1974ir} and subsequently we will refer to the $n_{x,\mu \nu} \in \mathds{Z}$ 
as the {\sl Villain variables}.

Finally we can address the topological charge $Q$, and as already announced in this section we now switch 
to d = 2 dimensions for this discussion. Similar to the gauge action in (\ref{SG_lat}) we define it 
in analogy to the continuum form (\ref{SG_cont}) as a sum over all lattice sites 
\begin{equation}
Q \; = \; \frac{1}{2 \pi} \sum_x F_{x,12} \; = \; \frac{1}{2 \pi} \sum_x \big((\textrm{d}\, a)_{x,12} \, + \, 2\pi \, n_{x,12} \big)  
\; = \; \sum_x  n_{x,12} \; \in \; \mathds{Z} \; .
\end{equation}
In the last sum we used the fact that the sum of the exterior derivative $(\textrm{d}\, a)_{x,12}$ over a lattice without 
boundaries vanishes (note that we use periodic boundary conditions which gives the lattice the topology of 
the 2-torus $\mathds{T}_2$). Thus the topological charge in 2 dimensions is simply the sum over the Villain 
variables $n_{x,12}$ and thus $Q$ is an integer by definition. 

We conclude the construction of our 
discretization of 2-d U(1) lattice gauge theory with a 
topological term by stating the path integral measure for the gauge fields which is given by (note that this 
is the form valid for arbitrary dimensions)
\begin{equation}
\sum_{\{ n \}} \; \int \! D[a] \; = \; \left[ \prod_{x} \prod_{\mu < \nu} \;\; \sum_{n_{x,\mu \nu} \in \mathds{Z}} \right] \; 
\left[ 
\prod_{x,\mu} \; \int_{-\pi}^{\pi} \frac{{d} a_{x,\mu}}{2 \pi} \right] \; .
\label{measure}
\end{equation}
The path integral measure is a sum over all configurations of the Villain variables $n_{x,\mu \nu} \in \mathds{Z}$ 
and an integral over all $a_{x,\mu} \in [-\pi, \pi]$ which for later convenience we normalize by $2\pi$. 

Putting things together we find the partition sum 
of 2-d U(1) lattice gauge theory with a topological 
term in the form,
\begin{equation}
Z\; = \; 	\sum_{\{ n \}} \; \int \! D[a] \;
\exp \left( - \, \frac{\beta}{2} \sum_{x} F_{x,12} \, F_{x,12}  \; + \; i \, \theta \, \sum_{x} n_{x,12} \right) \; ,
\label{Z_lat_2d}
\end{equation}
with $F_{12} \, = \, ( \textrm{d} \, a \; + \; 2 \pi \, n )_{x,\mu \nu} \; $.

\section{Testing the index theorem for the topological charge}

Having defined the new discretization and identified the form of the topological charge as the sum over 
the Villain variables, we now test properties of the topological charge and in particular explore if 
the Atiyah-Singer index theorem \cite{Atiyah} is implemented correctly. The index theorem is the key relation for 
the interaction of topological objects with fermions and a proper lattice discretization of the topological charge must 
suitably implement the index theorem. The corresponding study has been presented in 
\cite{Gattringer:2019yof} and we briefly summarize the main findings here. 

The index theorem links the difference of left- and right-handed zero modes of the Dirac operator to the topological charge,
\begin{equation}
Q \; = \; n_- \; - \; n_+ \; ,
\label{indextheorem}
\end{equation}
where $n_{+}$ ($n_-$) are the number of zero modes that are positive (negative) eigenstates of $\gamma_5$. Note that 
in 2 dimensions the so-called  {\sl vanishing theorem} in addition states that only one of the two numbers 
$n_+, n_-$ can be different from zero \cite{vanishing1,vanishing2,vanishing3}. 

For our analysis of the index theorem we need a chiral version of the lattice Dirac operator $D^{ov}$ 
and thus use the overlap operator \cite{overlap1,overlap2} defined as,
\begin{equation}
D^{ov} \; = \; \mathds{1} + A(\gamma_5 A \gamma_5 A)^{-1/2} \; \; \mbox{with} \; \; A = D^w - \mathds{1} \; ,
\label{DOV}
\end{equation} 
where in 2-d $\gamma_5 \equiv \sigma_3$ and $D^w$ is the Wilson-Dirac operator 
\begin{equation}
D^w_{\;x,y} \; = \; 2 \, \mathds{1}  \, \delta_{x,y} \, - \,
\sum_{\mu = 1}^2 \left[ \frac{\mathds{1} \!-\! \sigma_\mu}{2} \, U_{x,\mu} \, \delta_{x+\hat{\mu},y}
+ \frac{\mathds{1} \!+\! \sigma_\mu}{2} \, U_{y,\mu}^{\, *} \, \delta_{x-\hat{\mu},y} \right].
\label{DW}
\end{equation}
Here $\sigma_\mu$ are the Pauli matrices and as before $U_{x,\mu} = e^{\, i \, a_{x,\mu}}$.

For any operator that obeys the Ginsparg-Wilson relation (which is the case for the overlap operator) we can write the
rhs.\ of the index theorem (\ref{indextheorem}) in the form 
\begin{equation}
n_- \, - \, n_+ \; = \;  \frac{1}{2} \mbox{Tr} \big[ \gamma_5 D^{ov} \big] \; \equiv \;  Q_F \; ,
\label{QF}
\end{equation}
which constitutes the so-called fermionic definition of the topological charge \cite{indexlattice} that here
will be denoted as $Q_F$. 

The definition of the topological charge as the sum of the Villain variables, which we discussed in the previous section, 
will be referred to as $Q_V$, i.e.,
\begin{equation}
Q_V \; = \; \sum_{x} n_{x,12} \; ,
\label{QV}
\end{equation}
and we consider the index theorem correctly implemented when $Q_V = Q_F$ in the continuum limit. 

In addition to $Q_V$ and $Q_F$ we study yet another definition of the topological charge 
which can be constructed for our
formulation as follows: We may sum up the Villain variables on every plaquette and in this way identify
a summed Boltzmann factor $B[a]_{\beta, \theta}$ that is a functional of only the configurations of the fields 
$a_{x,\mu} \in  [-\pi, \pi]$ and depends on the parameters $\beta$ and $\theta$. Using this we can write
the partition function of pure gauge theory in the following form:
\begin{equation}
Z \; = \;  \int \! \! D[a] \;  B[a]_{\beta, \theta} \quad \mbox{with} \quad B[a]_{\beta, \theta} \; = \; 
\prod_x \; \sum_{n_{x,12} \in \mathds{Z}}
e^{\, - \, \frac{\beta}{2}\big((\textrm{\scriptsize d}\, a)_{x,12} + 2 \pi \, n_{x,12}\big)^2} \; e^{ \, i \, \theta n_{x,12}} \; .
\label{Bsummed}
\end{equation} 
It is rather straightforward to implement a Monte Carlo update\footnote{Note that we here talk 
about simulations at $\theta = 0$ that are free of the complex action problem. For
simulations at $\theta \neq 0$ see the next section.} with the Boltzmann factor $B[a]_{\beta, \theta}$,
since a change of some field variable $a_{x,\mu}$ only affects the 2 plaquettes that contain the link $(x,\mu)$. 
Only for these plaquettes the Boltzmann factor needs to be evaluated in a Metropolis step and  
the corresponding fast converging sums over the $n_{x,\mu \nu}$ can be computed with high accuracy 
at a reasonable numerical cost.

A topological charge for the discretization based on the formulation (\ref{Bsummed}) with summed Boltzmann 
factors can be obtained as the derivative of $\ln B[a]_{\beta, \theta}$ with respect to $\theta$ 
evaluated at $\theta = 0$. We refer to this definition of the topological charge as $Q_S$, and find
\begin{equation}
Q_S \; = \;   - i \, \frac{\partial}{\partial \theta} \ln B[a]_{\beta, \theta} \, \Bigg|_{\theta = 0} = \quad
\sum_x \;
\frac{\sum_{\,n_{x,12}}   \; e^{\; - \, \frac{\beta}{2} \big((\textrm{\scriptsize d} \, a)_{x,12} \; + \;  
2 \pi \, n_{x,12}\big)^2 } \; n_{x,12} }
{ \sum_{\,n_{x,12}}  \; e^{\; - \, \frac{\beta}{2} \big((\textrm{\scriptsize d} \, a)_{x,12} \; + \; 2 \pi \, n_{x,12}\big)^2 } } \; .
\label{QS}
\end{equation}
While the definition $Q_V$ in Eq.~(\ref{QV}) depends only on the Villain variables $n_{x,12}$, the definition $Q_S$ 
is a functional of only the $a_{x,\mu}$. Furthermore, is is obvious that $Q_S$ is not an integer and part of the 
numerical analysis below will be devoted to establishing that $Q_S$ becomes concentrated on integers when we 
approach the continuum limit. 

For the numerical analysis of the topological charge definitions and the index theorem we use quenched simulations with the Villain action on lattices with sizes ranging from $V = 8^2$ to $V = 24^2$ 
with statistics of typically $10^5$ configurations. The algorithm we use is a local Monte Carlo step that alternates
sweeps of updates of the $a_{x,\mu}$ and of the Villain variables $n_{x,12}$. In addition we perform simulations with the same parameters
also for the version with the summed Boltzmann factors as discussed above. The continuum limit is approached by 
sending both $\beta \rightarrow \infty$ and $V \rightarrow \infty$ by keeping the dimensionless ratio   
$R = V/\beta$ fixed, i.e., we approach the continuum limit with a fixed physical volume. 
More specifically we use three different values of $R$, $R = 32$, $R=64$ and $R = 128$, 
which correspond to continuum limits at different physical volumes.

\begin{figure}[t]
\begin{center}
   \includegraphics[width=0.48\linewidth,clip]{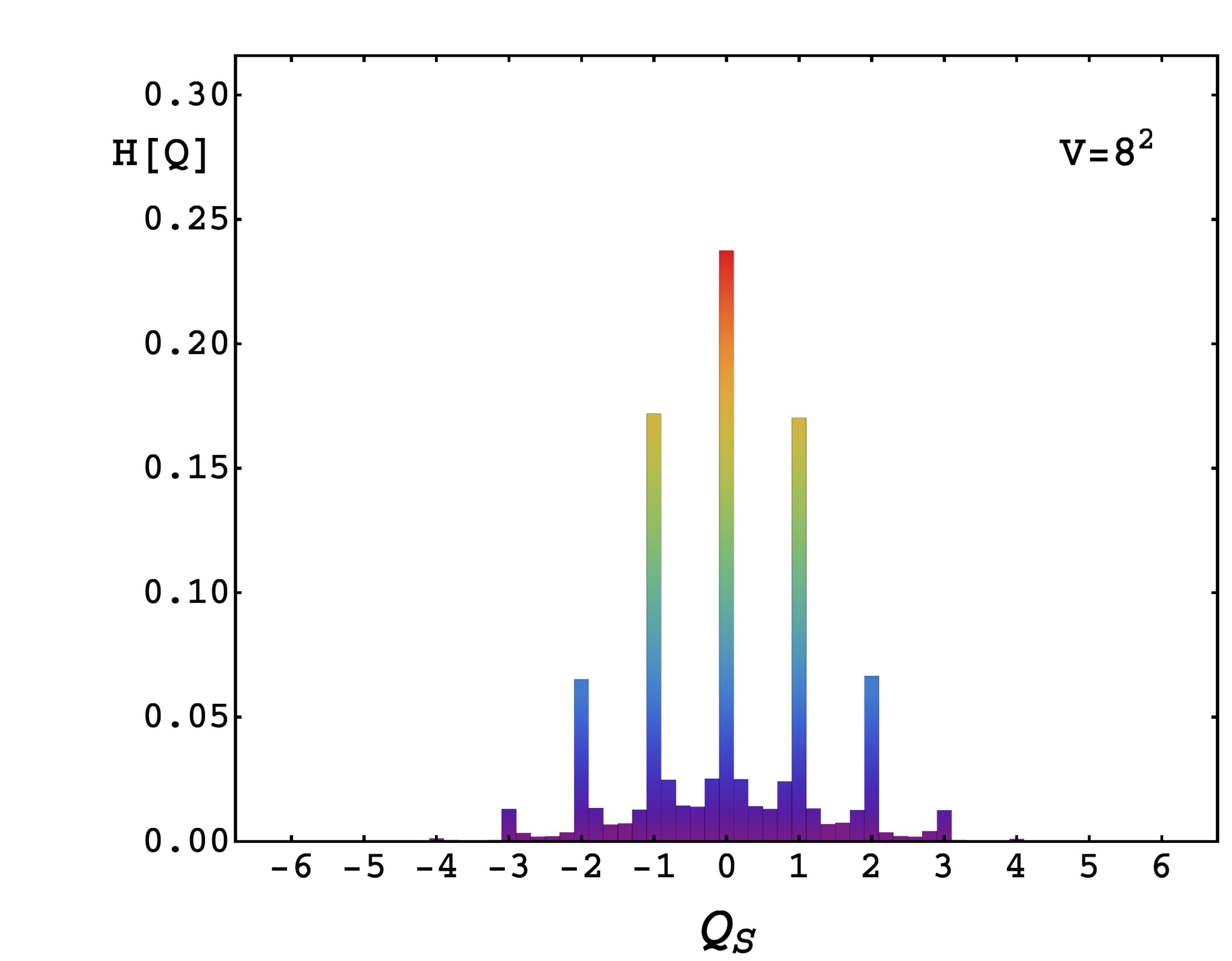}
   \includegraphics[width=0.48\linewidth,clip]{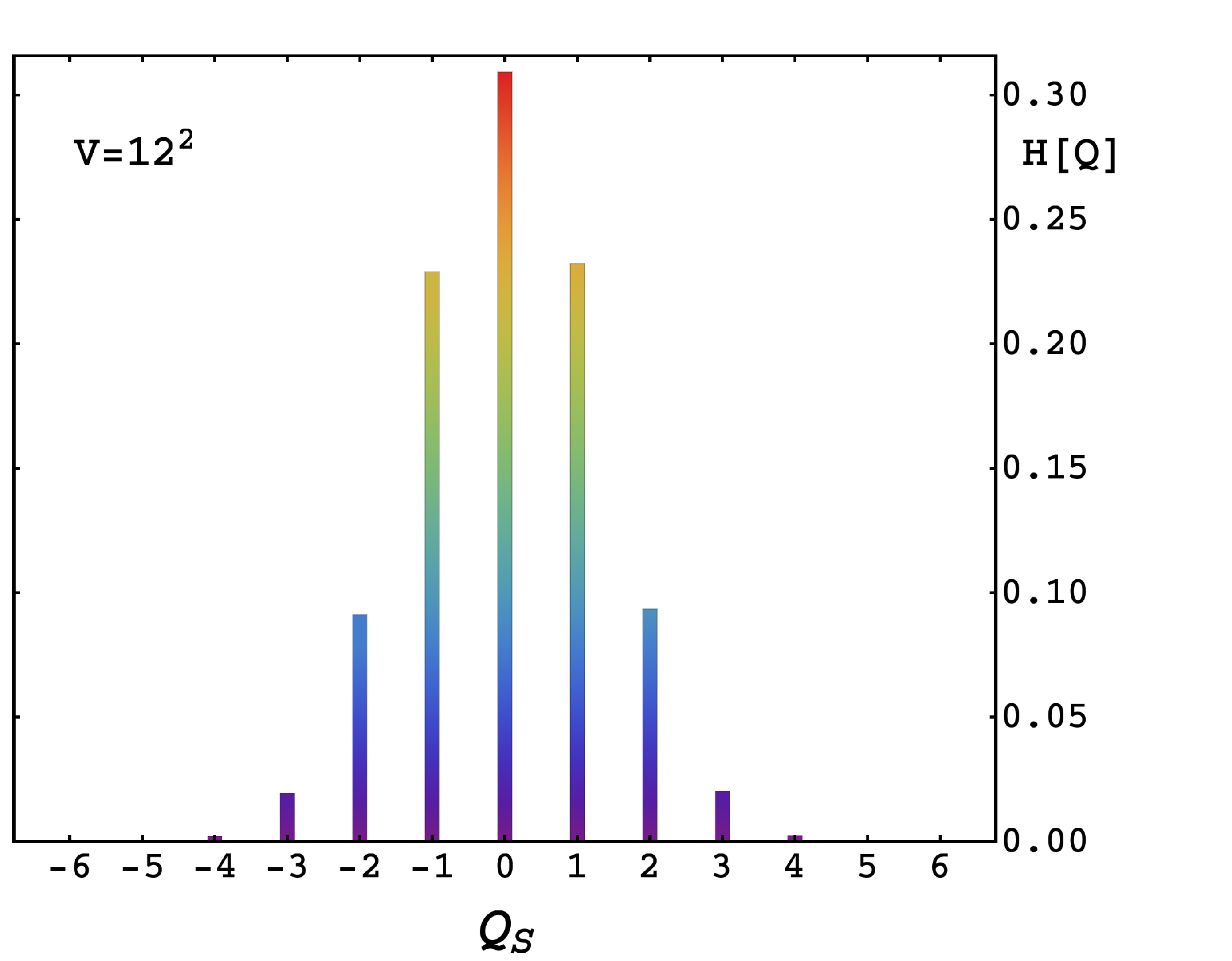}
\end{center}
\vspace{-5mm}
\caption{Distribution of the topological charge $Q_S$. In the lhs.~plot we show histograms for 
$\beta = 1.0$ with  $V = 8 \times 8$, while the rhs.\ data are for $\beta = 2.25, V = 12 \times 12$.
(Figure from \cite{Gattringer:2019yof}.)}
\label{fig:Qhisto}
\vskip5mm
\begin{center}
   \includegraphics[width=70mm,clip]{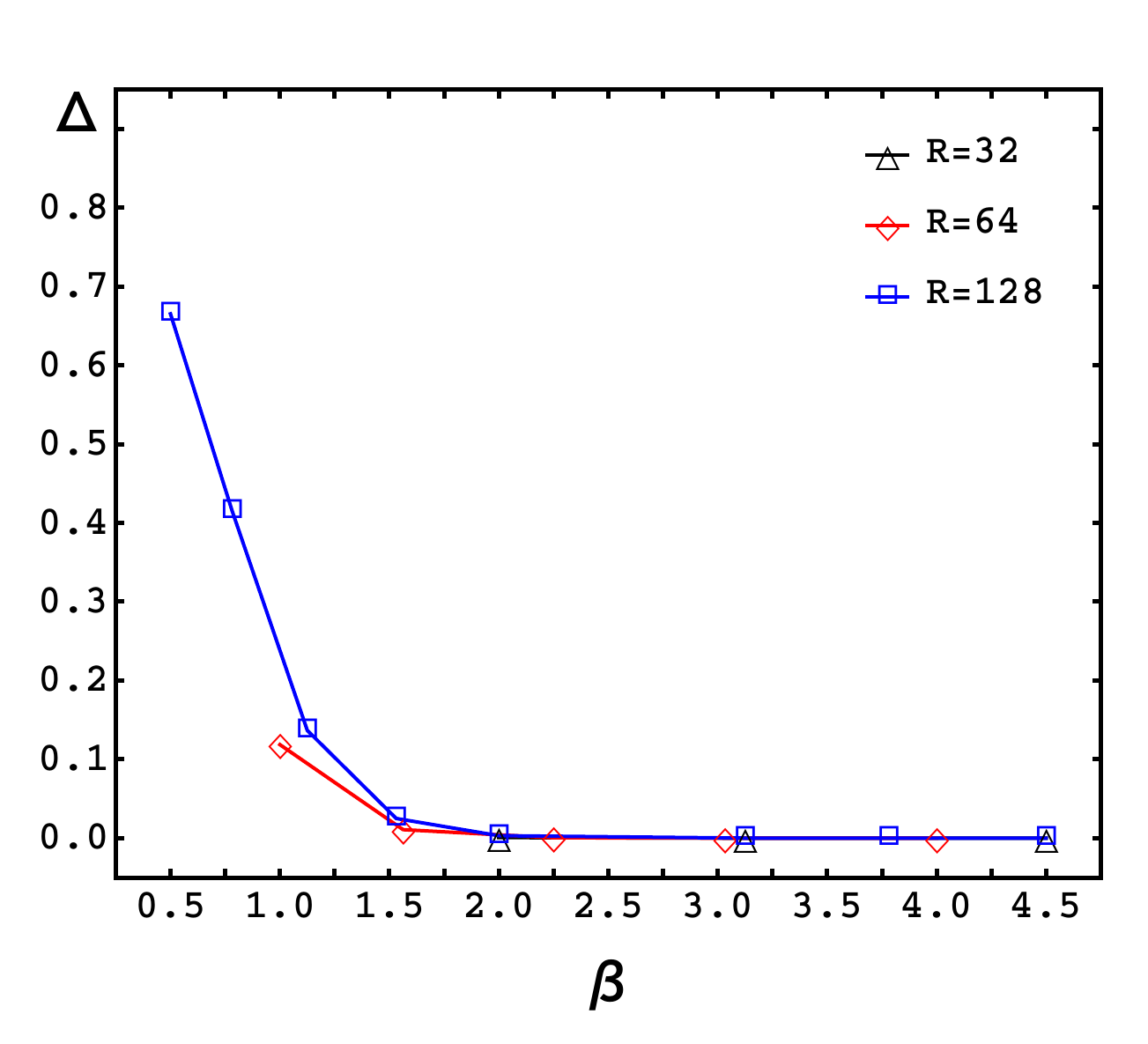} 
   \hspace{-4mm}
   \includegraphics[width=70mm,clip]{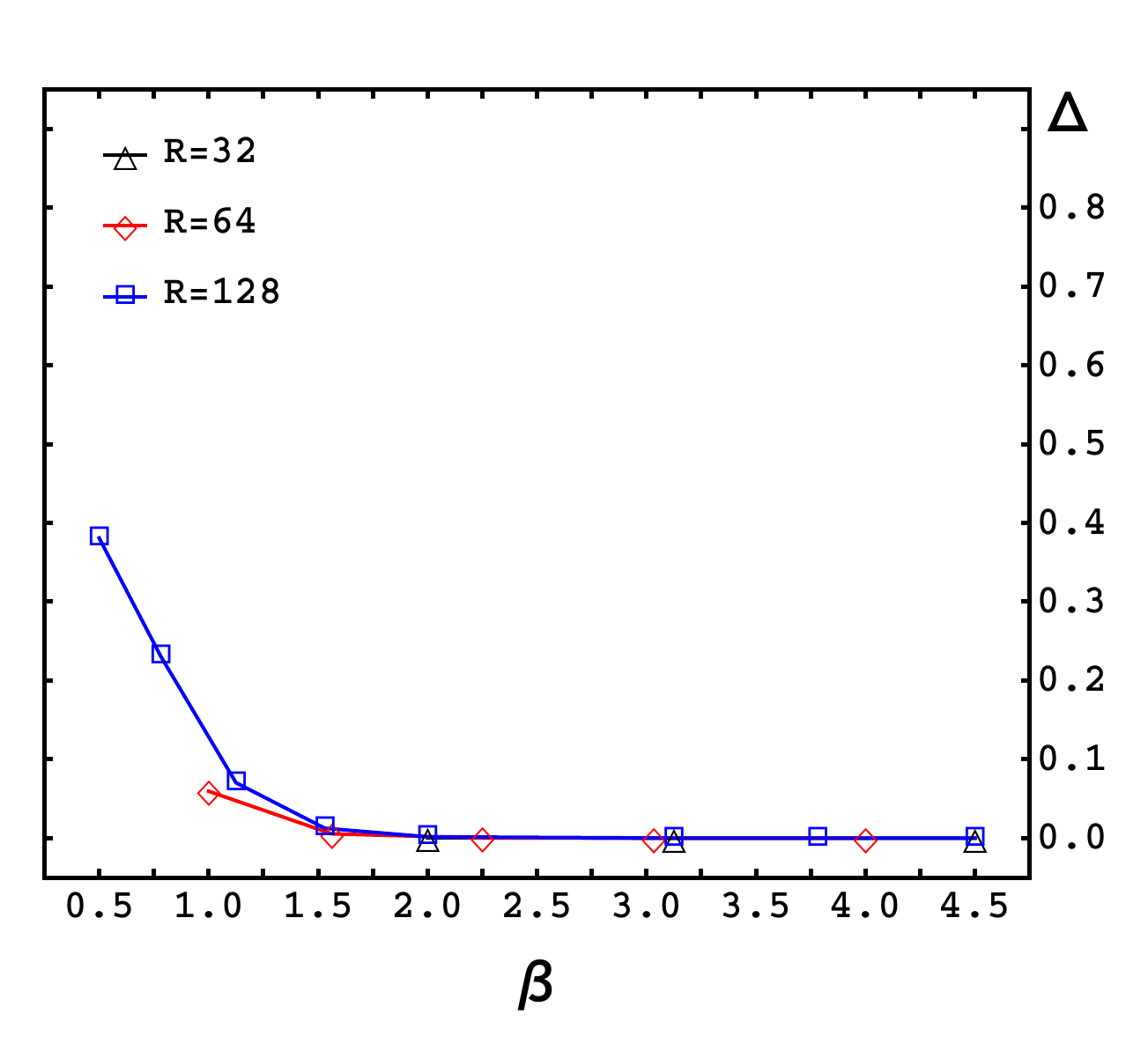}  
\end{center}  
\vspace{-5mm}
\caption{The fraction $\Delta$ of configurations with a mismatch of $Q_V$ and $Q_F$ (lhs.~plot) and a 
mismatch of $Q_S$ and $Q_F$ (rhs.). We show $\Delta$ as a function of $\beta$ and provide results for all 
ensembles. (Figure from \cite{Gattringer:2019yof}.)} 
\label{fig:mismatch}
\end{figure}

We begin our numerical analysis with studying the properties of the definition $Q_S$ of the topological charge.
It was already pointed out that $Q_S$ is not an integer but is expected to become concentrated on integers
when the continuum limit is approached. In Fig.~\ref{fig:Qhisto} we show histograms for the distribution of the 
values of $Q_S$ from our $R = 64$ ensembles. In the lhs.\ plot we show the distribution for 
$\beta = 1.0$ with $V = 8 \times 8$, while the rhs.\ plot is closer to the continuum limit with $\beta = 2.25$ and
$V = 12 \times 12$. While for the $\beta = 1.0$ configurations we still find some configurations with values
of $Q$ outside the bins at the integers, the rhs.\ plot for $\beta = 2.25$ shows that all histogram entries have 
become concentrated on integers as expected.  

Let us now come to the assessment of the index theorem for our quenched ensembles. For each configuration 
we exactly solve the eigenvalue problem for the Hermitian matrix $\gamma_5 \, A$  
(see Eq.~(\ref{DOV})) using linear algebra packages. From that we construct the overlap operator using its
spectral representation and subsequently determine the zero modes that we need for the definition of $Q_F$.
In addition we compute $Q_V$ and $Q_S$ for each configuration and compare these values to $Q_F$. 
To quantify the violation of the index theorem based on $Q_V$ we determine the fraction $\Delta$ 
of configurations where $Q_V$ and $Q_F$ give different results. For $Q_S$ we first round to the nearest integer 
before we compute the fraction $\Delta$ for the mismatch of $Q_S$ and $Q_F$. 

In the lhs.~plot of Fig.~\ref{fig:mismatch} we show our results for the mismatch $\Delta$ between $Q_V$ and $Q_F$, 
while the rhs.\ shows $\Delta$ for $Q_S$ and $Q_F$. We plot $\Delta$ as a function of the inverse gauge coupling 
$\beta$ and show the results for all values of the ratio $R$ we considered. For small values of $\beta$ we observe 
a sizable fraction of mismatch, but it is obvious that $\Delta$ decreases quickly as $\beta$ is increased, and above
$\beta \sim 2.5$ the values of $\Delta$ are essentially 0 for both definitions $Q_V$ and $Q_S$ and all three ratios $R$. Thus we
conclude that the index theorem is obeyed in the continuum limit for both $Q_V$ and $Q_S$.

\section{The 2-d U(1) gauge Higgs model with topological term and its worldline/worldsheet representation}

We now come to the discussion of systems where our gauge action with topological term is coupled to
matter fields. More specifically we consider the 2-d U(1) gauge Higgs model with two flavors and a topological term,
where we describe the gauge degrees of freedom with the formalism developed in Section 2. 
The partition sum of the gauge Higgs model reads 
\begin{equation}
Z \; = \; \sum_{\{n\}} \, \int \!\! D[a] \, \int \!\! D[\varphi] \, \int \!\! D[\chi] \; 
e^{\; - \, S_g[a,n] \; + \; i \, \theta \, Q[n] \; - \; S_h[a,\varphi,\chi] } \; ,
\label{Z_GH}
\end{equation}
with the gauge action, the topological charge and the Higgs field action given by
\begin{eqnarray}
S_g[a,n] & = & \frac{\beta}{2} \sum_x \big( (\textrm{d}\,a)_{x,12} + 2\pi \, n_{x,12} \big)^2 \qquad , \qquad 
Q[n] \; = \; \sum_x n_{x,12} \; ,
\label{SG_gh}
\\
S_h[a,\varphi,\chi] & \!\! = \!\! & \sum_{x\in\Lambda}\Big[ M \big( |\varphi_x|^2 + |\chi_x|^2\big) + 
\lambda \big( |\varphi_x|^4 + 2g |\varphi_x|^2 |\chi_x|^2 + |\chi_x|^4\big)  
\nonumber \\
& & \hspace{40mm} - \sum_{\mu=1}^{2} 
\big(\varphi_{x}^{\ast} U_{x,\mu} \varphi_{x+\hat{\mu}}+\chi_{x}^{\ast} U_{x,\mu} \chi_{x+\hat{\mu}} \, + \, c.c.\big) \Big] \; .
\label{SH_gh}
\end{eqnarray}
The two flavors of scalars are described by $\phi_x, \chi_x \in \mathds{C}$ and couple to the gauge degrees of freedom.\ 
via the gauge links  $U_{x,\mu} = e^{\, i a_{x,\mu}}$ as discussed in Section 2. 
$M = 4 + m^2$ combines the constant term of the discretization of the Laplace operator with the mass parameter 
$m^2$ which we choose equal for the two flavors. $\lambda$ is the coupling for the quartic self interaction of the two 
flavors where we also introduced a mixing parameter $g \geq -1$. 
The corresponding path integral measures are defined as 
\begin{equation}
	\sum_{\{n\}} \, = \, \prod_{x} \; \sum_{n_{x,12} \in \mathds{Z}} \; , \; 
	\int \!\! D[a] \, = \, \prod_{x,\mu} \int_{-\pi}^{\pi} \!\! \frac{da_{x,\mu}}{2\pi} \; , \;
	\int \!\! D[\varphi] \, = \, \prod_x \int_{\mathds{C}} \!\! \frac{d\varphi_x}{2\pi} \; , \;
	\int \!\! D[\chi] \, = \, \prod_x \int_{\mathds{C}} \!\! \frac{d\chi_x}{2\pi}\;  .
\label{measures}	
\end{equation}
We remark that it is straightforward to reduce the model to only a single flavor by dropping all terms that contain 
$\chi_x$.

The U(1) gauge Higgs models with a topological term are particularly interesting because they are related to half-integer spin chains at $\theta = \pi$, and obey interesting constraints on their IR physics coming from 't Hooft anomaly matching \cite{Gaiotto:2017yup,Sulejmanpasic:2018upi,Tanizaki:2018xto}. Charge conjugation is implemented in our lattice 
discretization by the following transformations,
\begin{equation}
a_{x,\mu} \; \rightarrow \; - a_{x,\mu}\; , \; \; 
n_{x,12} \; \rightarrow \; - n_{x,12} \; , \; \; 
\varphi_{x} \; \rightarrow \; \varphi_x^* \; , \; \; 
\chi_{x} \; \rightarrow \; \chi_x^* \; .
\label{chargeconjugation}
\end{equation}
While the gauge and Higgs field action are invariant under charge conjugation the topological charge $Q[n]$ 
changes its sign. Since in our formulation $Q[n]$ is an integer and enters the partition sum 
(\ref{Z_GH}) in the form $e^{ \, i \, \theta Q[n]}$ it is obvious that the two values $\theta = 0$ and $\theta = \pi$ 
are the two points where the system is invariant under charge conjugation. $\theta = 0$ is the trivial 
implementation, but for $\theta = \pi$ the 
$\mathds{Z}_2$ charge conjugation symmetry (\ref{chargeconjugation}) 
can be broken spontaneously. A study of the corresponding physics in the one-flavor model has been 
presented in \cite{Gattringer:2018dlw,Goschl:2018uma} and in the next section we will discuss first results for 
the two-flavor case. 

Before we can discuss numerical results we need to address the complex action problem of the partition sum (\ref{Z_GH}).
It is obvious that for $\theta \neq 0$ the Boltzmann factor is complex and cannot be used as a probability in a Monte
Carlo simulation. However, also with the our gauge field discretization 
the model has an exact mapping \cite{Gattringer:2018dlw,Goschl:2018uma,Sulejmanpasic:2019ytl} 
to a real and positive representation in terms of worldlines and worldsheets\footnote{For the worldline/worldsheet representation in the compact formulation see \cite{Kloiber:2014dfa,Gattringer:2015baa}.}. 

The worldline representation uses link-based flux variables $j_{x,\mu}, k_{x,\mu} \in \mathds{Z}$ 
for describing the Higgs field degrees of freedom. The gauge degrees of freedom are described 
by plaquette occupation numbers $p_x \in \mathds{Z}$, which in 2-d we can label by the coordinate
$x$ of the site in the lower left corner of the plaquette.  In the worldline/worldsheet form of the
partition function we sum over all configurations of these variables and introduce
\begin{equation}
\sum_{\{p\}} \; = \;  \prod_x \sum_{p_x \in \mathds{Z}} \quad , \quad  
\sum_{\{j,k\}} \; = \;  \prod_{x,\mu} \sum_{j_{x,\mu}\in\mathds {Z}} \; \sum_{k_{x,\mu}\in\mathds{Z}} \; .
\end{equation}
The sums over the fluxes and plaquette occupation numbers are subject to constraints and each admissible 
configuration comes with real and positive weight factors, such that the partition function is given by
(we here denote the Kronecker delta with 
$\delta(j) \equiv \delta_{j,0}$)
\begin{eqnarray}
\hspace*{-8mm} && \hspace*{25mm} Z  \; =  \;  \sum_{\{p\}}  \, W_g[p]  \; \sum_{\{j,k\}} W_h[j,k] \; \times  
\label{dualZ} \\
\hspace*{-8mm} &&
\prod_x  \delta\!\left(\!\vec{\nabla}\cdot \vec{j}_x\! \right) \delta\!\left(\!\vec{\nabla}\cdot \vec{k}_x\! \right) 
\delta\!\left(j_{x,1}+k_{x,1}+p_{x}-p_{x-\hat{2}}\right)  \delta\!\left(j_{x,2}+k_{x,2}-p_{x}+p_{x-\hat{1}}\right) .
\nonumber
\end{eqnarray}
The weight factor $W_g[p]$ for the configurations of the plaquette occupation numbers depends on the 
inverse gauge coupling $\beta$ and the topological angle $\theta$,
\begin{equation}
W_g[p] \; = \; \left( 2\pi \beta \right)^{-\frac{V}{2}} 
\prod_x e^{\, - \frac{1}{2 \beta} \big(p_x + \frac{\theta}{2\pi}\big)^2} \; .
\label{W_G}
\end{equation}

The plaquette occupation numbers are Gauss-distributed and the topological angle has the effect of shifting the 
center of the Gaussian. We stress that $\theta = 0$ and $\theta = \pi$ are symmetry points of the distribution 
corresponding to $p_x \rightarrow - p_x$ for $\theta = 0$ and $p_x \rightarrow - p_x - 1$ for $\theta = \pi$, which 
together with $j_{x,\mu} \rightarrow - j_{x,\mu}, k_{x,\mu} \rightarrow - k_{x,\mu}$ implement the trivial ($\theta = 0$) 
and the non-trivial ($\theta = \pi$) charge conjugation symmetry in the worldline/worldsheet representation. 

For the configurations of the fluxes $j_{x,\mu}$ and $k_{x,\mu}$, which describe the Higgs fields in the worldline
representation, the weight factor $W_h[j,k]$ is given by
\begin{eqnarray}
\label{equ:dual_WH}
&& W_h[j,k] =  \sum_{\{\overline{j},\overline{k}\}} \, \prod_{x,\mu} \, 
\frac{1}{\left(|j_{x,\mu}|\!+\!\overline{j}_{x,\mu}\right)! \ \overline{j}_{x,\mu}!} \, 
\frac{1}{\left(|k_{x,\mu}|\!+\!\overline{k}_{x,\mu}\right)! \ \overline{k}_{x,\mu}!} \; \prod_x  I\left(f_x,g_x\right) 
\quad \mbox{with} 
\label{weightfactors}\\
&& 
I\!\left(f_x,g_x\right)  =  \int_{0}^{\infty} \!\!\!\!\! dr \int_{0}^{\infty} \!\!\!\!\! ds \; r^{\,f_x+1} \; s^{\,g_x+1} \; 
e^{\, - \, M (r^2 + s^2) \; - \, \lambda (r^4 + 2 g r^2 s^2 + s^4) }  \; ,
\nonumber \\
&& 
f_x = \sum_{\mu}
\left[ |j_{x,\mu}| + |j_{x-\hat{\mu},\mu}| + 2\left( \overline{j}_{x,\mu} + \overline{j}_{x-\hat{\mu},\mu} \right) \right] \; , \; \;
g_x =\ \sum_{\mu}
\left[ |k_{x,\mu}| + |k_{x-\hat{\mu},\mu}| + 2\left( \overline{k}_{x,\mu} + \overline{k}_{x-\hat{\mu},\mu} \right) \right] \; .
\nonumber 
\end{eqnarray}
These weight factors are sums over unconstrained auxiliary link-based variables 
$\overline{j}_{x,\mu}, \overline{k}_{x,\mu} \in \mathds{N}_0$ with  $\sum_{\{\overline{j},\overline{k}\}} \equiv 
\prod_{x,\mu} \sum_{\overline{j}_{x,\mu}\in\mathds {N}_0} \sum_{\overline{k}_{x,\mu}\in\mathds{N}_0}$ that 
may be treated with standard Monte Carlo methods. The integrals $I(f,g)$ obviously take into account 
the on-site terms of the action (\ref{SH_gh}). For the simulation purposes these are numerically 
pre-computed and stored for sufficiently many values of the integers $f_x$ and $g_x$ that are combinations 
of the flux and auxiliary variables for the two flavors.

Finally we need to discuss the constraints that have to be obeyed by admissible configurations of the flux variables 
$j_{x,\mu}, k_{x,\mu}$ and the plaquette occupation numbers $p_x$. In the partition sum  (\ref{dualZ}) the constraints 
are implemented with products of Kronecker deltas and we can distinguish two types of constraints. The 
first set of constraints are zero divergence constraints for the flux variables which have the form
\begin{equation}
\vec{\nabla}\cdot \vec{j}_x \equiv \sum_{\mu} [ j_{x,\mu} - j_{x-\hat{\mu},\mu}] = 0 \quad \forall x \; , \quad
\vec{\nabla}\cdot \vec{k}_x \equiv \sum_{\mu} [ k_{x,\mu} - k_{x-\hat{\mu},\mu}] = 0 \quad \forall x \; .
\end{equation}
These constraints require flux conservation of the $j_{x,\mu}$ and the $k_{x,\mu}$ at every site, such that in
admissible configurations of the flux variables these must form closed loops. 
The second constraint comes from gauge invariance and for every link of the lattice requires that the total 
flux from the two plaquette occupation numbers attached to that link and the combined $j$- and $k$-flux vanishes. 

The worldline/worldsheet representation (\ref{dualZ}),
(\ref{W_G}), (\ref{weightfactors}) can be obtained by using high temperature expansion techniques
for the gauge- and Higgs field Boltzmann factors, essentially following the strategy developed for the 4-d case 
\cite{Mercado:2013yta,Schmidt:2012uy,Mercado:2013ola,Mercado:2013jma}. For treating the topological term
in the mapping to the worldline/worldsheet representation one uses Poisson resummation, which gives rise
to the gauge field weight factors (\ref{W_G}). For details we refer to 
\cite{Gattringer:2018dlw,Goschl:2018uma,Sulejmanpasic:2019ytl}. 

Let us finally discuss observables in the worldline/worldsheet representation. Here we focus on simple 
bulk observables and their moments which can be obtained as derivatives of $\ln Z$ with respect to the parameters.
When the worldline/worldsheet representation is used for $Z$ the derivatives generate the observables directly
in terms of the flux variables and plaquette occupation numbers.  

First and second derivatives of $\ln Z$ with respect to the topological angle $\theta$ 
generate the expectation values of the topological charge density $\!\langle q \rangle$ with 
$q = Q/V$, as well as
the topological susceptibility $\chi_t$. A few lines of algebra lead to (note that for $\chi_t$ an 
irrelevant additive constant was dropped),
\begin{eqnarray}
\hspace*{-8mm} && \langle q \rangle   =  - \frac{1}{V} \frac{\partial}{\partial\theta} \ln Z \; = \;  
\frac{1}{V} \left \langle \frac{1}{2\pi\beta}\sum_x \! \left[  p_x + \frac{\theta}{2\pi} \! \right]\! \right\rangle \; , 
\label{qdual}
\\
\hspace*{-8mm} && \chi_t  =  \frac{1}{V} \frac{\partial^2}{\partial\theta^2} \ln Z \; = \; 
\frac{1}{V} \! \left[ \left \langle \!\! \left(\!\frac{1}{2\pi\beta} \! \sum_x \! 
\left[  p_x + \frac{\theta}{2\pi} \! \right]\! \right)^{\!\!\!2}\right\rangle - 
\left \langle \! \frac{1}{2\pi\beta}\!\sum_x \! \left[  p_x + \frac{\theta}{2\pi} \! \right]\! \right\rangle^{\!\!\!2} \right] \; .
\nonumber
\end{eqnarray}

In the two flavor model the quartic term in the Higgs action (\ref{SH_gh}) allows for different types of flavor symmetries
depending on the couplings $\lambda$ and $g$. Thus an interesting question is to analyze possible flavor symmetry 
breaking and for such an analysis we introduce the flavor symmetry breaking parameter 
$c \equiv |\varphi_x|^2 - |\rho_x|^2$ and the corresponding susceptibility $\chi_c$. To generate these observables from 
$\ln Z$ we allow for two different mass parameters $M_\varphi$ and $M_\chi$, compute the necessary first 
and second derivatives with respect to $M_\varphi$ and $M_\chi$ and subsequently set $M_\varphi = M_\chi = M$.
Again we arrive at the corresponding worldline/worldsheet representation after a few lines of algebra, 
\begin{equation}
\langle |c| \rangle  \; = \; \frac{1}{V} \left\langle \sum_x \frac{| I(f_x + 2,g_x) - I(f_x,g_x+2) |}{I(f_x,g_x)} \right\rangle \; ,
\end{equation}
and a similar expression in terms of moments of $I(f_x,g_x)$ can be obtained for the corresponding 
flavor breaking symmetry susceptibility $\chi_c$. Finally we will also consider the Binder cumulant $U_t$
for the topologogical charge and the Binder cumulant $U_c$ for the flavor breaking observable,
\begin{equation}
	U_t \; = \; 1 \, - \, \frac{\langle q^4\rangle}{3\langle q^2\rangle^2} \quad , \qquad 
	U_c \; = \; 1 \, - \, \frac{\langle c^4\rangle}{3\langle c^2\rangle^2} \; ,
\end{equation}
and for the necessary observables we may obtain the corresponding worldline/worldsheet representations by 
suitable combinations of derivatives of $\ln Z$ with respect to the parameters.

\section{Numerical study of the 2-d gauge Higgs models at $\theta = \pi$}

Having discussed the 2-d U(1) gauge Higgs model in its worldline/worldsheet representation that overcomes the 
complex action problem, we now come to the discussion of selected preliminary numerical results\footnote{A publication 
with a complete presentation of all our results is in preparation \cite{WIP2}.} for the 2-flavor 
model at vacuum angle $\theta = \pi$. As we have already discussed, at $\theta = \pi$ the model implements charge
conjugation symmetry as a $\mathds{Z}_2$ symmetry in a non-trivial way. For the 1-flavor 2-d U(1) gauge Higgs model
we have shown that this symmetry can be broken spontaneously with the mass parameter $M$ being the 
coupling that drives the transition \cite{Gattringer:2018dlw,Goschl:2018uma} and the expectation 
value of the topological charge density $q$ as the corresponding order parameter. 
In the 2-flavor model the deformation parameter $g$ in the quartic 2-flavor self interaction term 
$\lambda \big( |\varphi_x|^4 + 2g |\varphi_x|^2 |\chi_x|^2 + |\chi_x|^4\big)$ in the Higgs action (\ref{SH_gh}) 
of the model allows one to control the symmetry of the model. While for $g = 1$ the symmetry is the full SO(3) 
flavor symmetry, for $g \neq 1$ the flavor symmetry is reduced to\footnote{Note that the global symmetry is always defined modulo the $U(1)$ gauge transformation.} $O(2)\subset SO(3)$. Thus we expect that $g$ is a relevant coupling for driving a spontaneous 
breaking of flavor symmetry, with the flavor breaking observable $c$ as the corresponding order parameter.
Indeed, a classical  analysis \cite{WIP2} indicates that in the $M$-$g$ phase diagram 
three different phases can be expected that are characterized by charge and flavor symmetry. 

The numerical analysis \cite{WIP2} will map out the details of the phase diagram and explore the nature of the 
critical lines. Here we present some preliminary results for the $M$-$g$ phase diagram and characterize the
phases with respect to their symmetries. 

Our Monte Carlo simulation is based on the worldline/worldsheet representation discussed in the previous section and 
uses different types of updates, which are combined to reduce autocorrelations. For the unconstrained dual variables 
$\overline{j}_{x,\mu}, \overline{k}_{x,\mu} \in \mathds{N}_0$ we simply use local Metropolis updates. The constrained 
variables $j_{x,\mu}, k_{x,\mu}, p_x \in \mathds{N}_0$ we update such that the constraints are automatically fulfilled. 
This is done locally by inserting a closed loop of $j$- or $k$-flux around a plaquette and changing the plaquette 
occupation number $p_x$ to compensate the loop-induced link flux. To reduce critical slowing down we add the surface worm algorithm presented in \cite{Mercado:2013yta}. For full ergodicity we include 
a second worm update which creates closed charge-neutral worms by jointly propagating a combination of both 
fluxes $j_{x,\mu}$ and $k_{x,\mu}$ through the lattice, thus allowing for the possibility of double loops 
for both
flavors jointly winding around the 
periodic boundaries. Additional decorrelation is achieved by explicitly using the $\mathbb{Z}_2$ symmetries as 
global updates as well as offering a global change of all plaquette occupation numbers. 

In our simulations we typically use $5\times 10^5$ equilibration sweeps before performing measurements. 
The accumulated statistics depends on the lattice size and the proximity to the critical lines and varies between 
$10^4$ and $10^7$ measurements which are decorrelated by $10$ to $100$ update sweeps. To be able to resolve 
the peaks of the susceptibility and Binder cumulant intersections with good precision we use a multiple 
histogram reweighting technique, where we perform simulations at around 5 parameter values in the critical region. 
The statistical errors of the simulations were computed
 with a Jackknife analysis. 

In order to determine the critical lines in the $M$-$g$ phase diagram we use horizontal ($g$ fixed, $M$ variable)
and vertical cuts ($g$ variable, $M$ fixed) and study observables as a function of the respective variable coupling.
The observables for locating the transition lines are the peak positions of the order parameter susceptibilities,
as well as the size-independent points of the Binder cumulants of the order parameters at the critical points. 

\begin{figure}[t]
\centering
	\includegraphics[width=0.5\textwidth,clip]{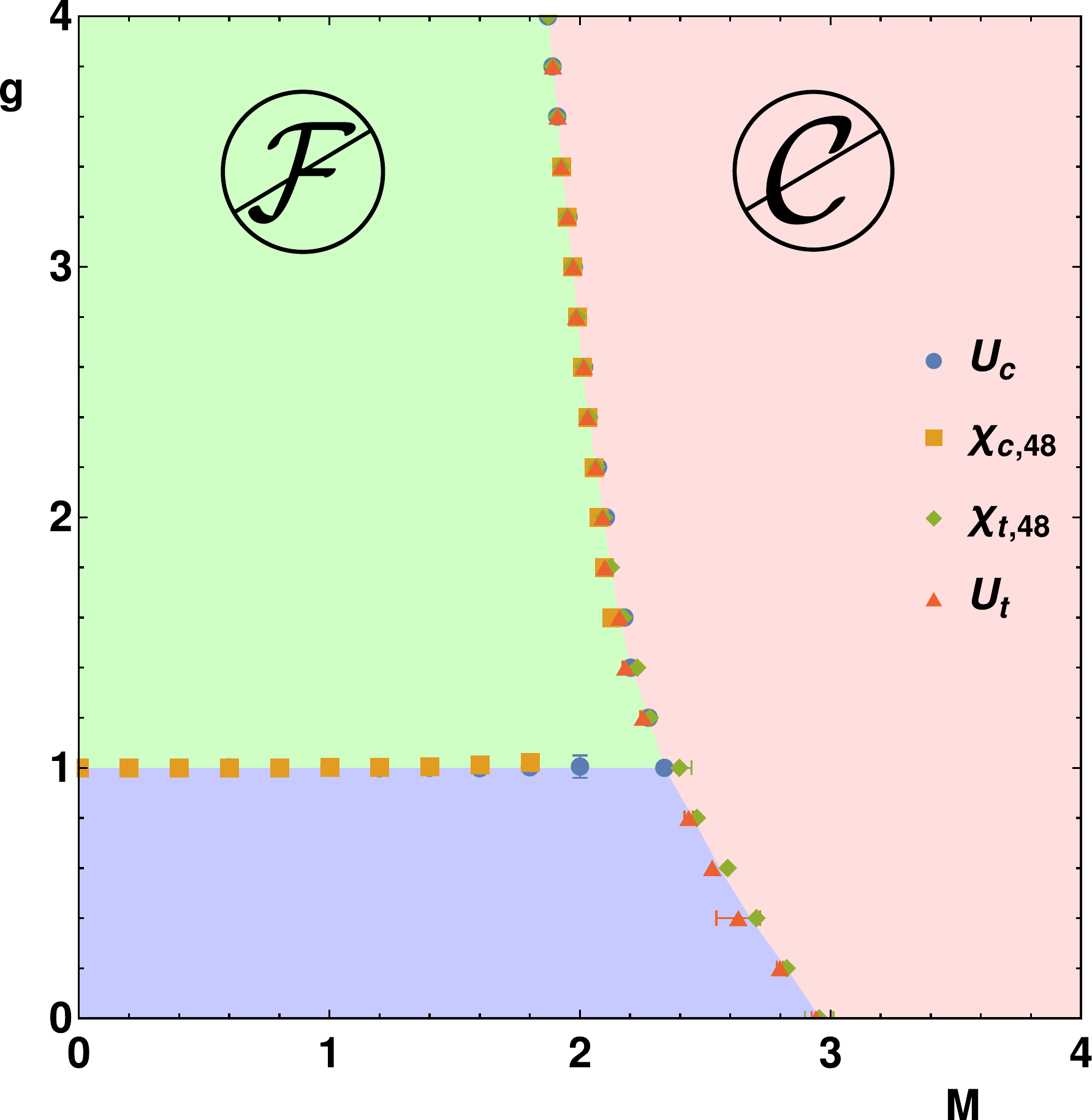} 
	\caption{Phase diagram of the 2-flavor U(1) gauge Higgs model at $\theta = \pi$ as a function of $M$ and $g$. 
	We show the peak locations of the flavor breaking susceptibility $\chi_c$ and of the topological susceptibility 
	$\chi_t$ from simulations on a $V=48^2$ lattice. In addition we show the intersection points of the corresponding 	
	Binder cumulants $U_c$ and $U_t$ using lattice sizes $V=32^2$ and $V=48^2$. In the top left phase 
	(green area in the plot) flavor symmetry is broken spontaneously, while in the phase on the rhs.\ of the diagram (red)  
	charge conjugation symmetry is broken. In the bottom left phase (blue area) both symmetries are intact.}
	\label{fig:phase_diagram}
\end{figure}

In Fig.~\ref{fig:phase_diagram} we show first results for the phase diagram as a function of the mass parameter $M$ and 
the deformation parameter $g$. The peak locations of the topological susceptibility $\chi_t$ and the flavor breaking 
susceptibility $\chi_c$ were measured on a $V=48^2$ lattice, while we determined the size independent point of the Binder 
cumulants $U_t$ and $U_c$ by taking the intersections of curves measured on $V=32^2$ and $V=48^2$ lattices. 
We find that the different determinations of the transition lines agree very well, but of course there will be 
residual finite volume corrections\footnote{Most notably there will be logarithmic corrections due to marginally irrelevant operators along the critical lines.} that will slightly shift the positions of the phase 
boundaries. For $g>1$ and small values of $M$, i.e., the green area in Fig.~\ref{fig:phase_diagram}, 
flavor symmetry is spontaneously broken, while for large values of $M$ (red area) the charge conjugation 
symmetry is broken spontaneously. For values $g<1$ and small values of $M$ (blue area) both symmetries are 
intact and the system is in a compact massless scalar phase.

\begin{figure}[t]
	\centering
	\includegraphics[width=0.97\textwidth,clip]{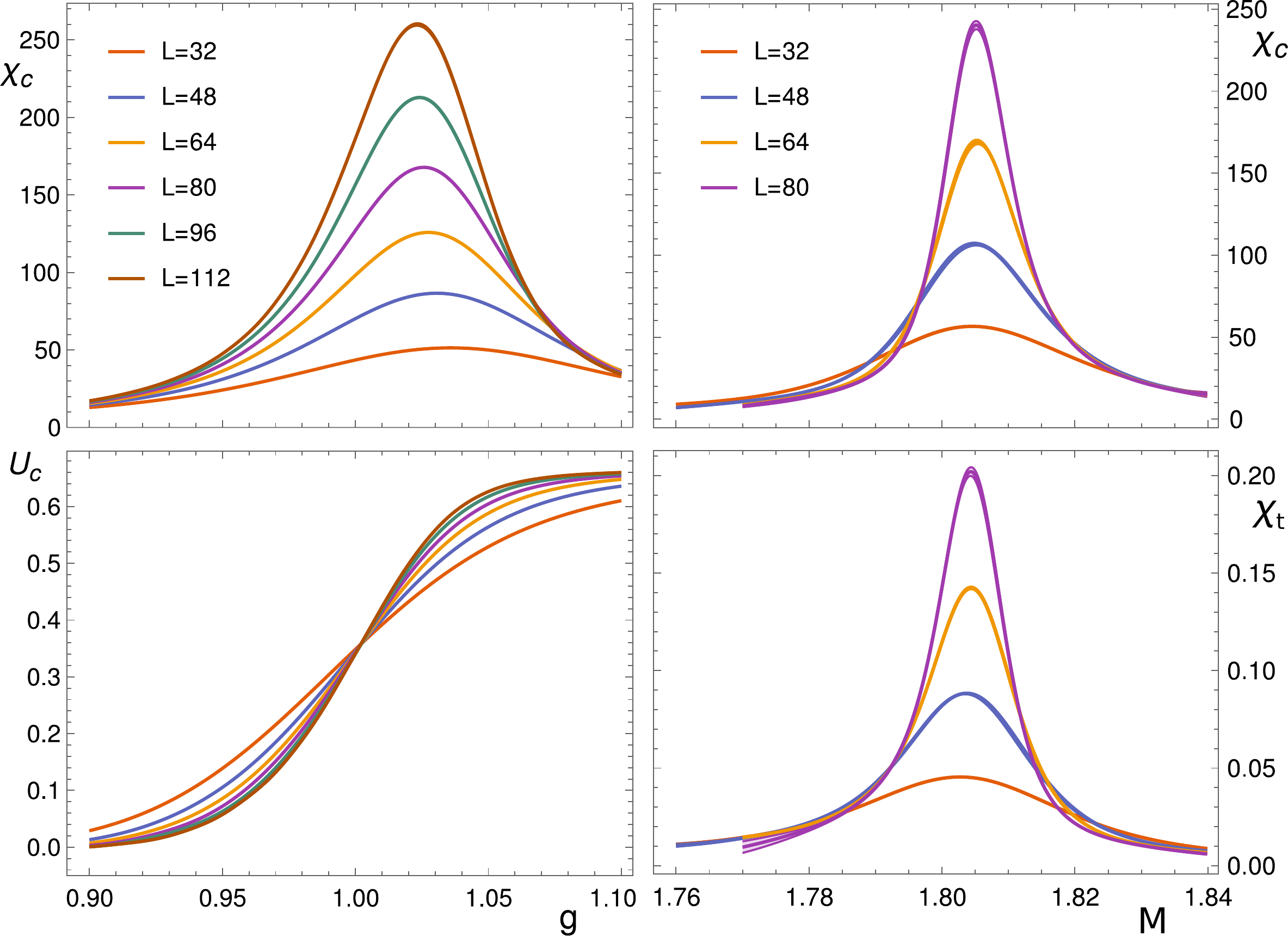} 
	\caption{Finite volume behavior of observables across critical lines. Lhs.: For a vertical cut at $M=1.8$ 
	we plot the flavor breaking susceptibility $\chi_c$ (lhs.\ top) and the corresponding Binder cumulant 
	$U_c$ (lhs.\ bottom) as a function of the  parameter $g$. Rhs.: For a horizontal cut at $g=5$ 
	we plot the flavor breaking susceptibility $\chi_c$ (rhs.\ top) and the topological susceptibility $X_t$ 
	(rhs.\ bottom) as a function of the mass parameter $M$.}
	\label{fig:critical_points}
\end{figure}

As already mentioned, the phase boundaries in Fig.~\ref{fig:phase_diagram} are still subject to finite size effects such that shifts of the critical lines are to be expected. In order to give a more detailed picture of our methodology, 
in Fig.~\ref{fig:critical_points} we explicitly demonstrate the analysis for two critical points in the phase diagram. 
On the left hand side we focus on the crossing of the critical line at fixed $M=1.8$ by varying the deformation 
parameter $g$, i.e., a vertical cut. We observe a spontaneous breaking of the flavor symmetry as a function of $g$
and show the flavor breaking susceptibility $\chi_c$ (lhs.\ top) and the flavor breaking Binder cumulant $U_c$ 
(lhs.\ bottom) for 
different volumes $V=L^2$. As expected, we observe peaks in the susceptibilities which slowly converge towards 
smaller values of $g$, giving an increasingly precise estimate for the critical point. The Binder cumulant $U_c$ plotted for 
different volumes shows the expected behavior of vanishing in the symmetric phase and approaching a value of $2/3$ in 
the broken phase. In between, at $g\approx 1$, there is a universal point where $U_c$ does not scale, providing an 
estimate for the critical point which has only very small finite size corrections. 
On the right hand side we show results for a horizontal cut at fixed $g=5$ where the phase boundary is crossed when 
varying the mass parameter $M$. The flavor symmetry is spontaneously broken for small values of $M$,
while for large values of $M$ charge conjugation symmetry is broken. We show results for both susceptibilities, 
i.e., the flavor breaking susceptibility $\chi_c$ (rhs.\ top) and the topological susceptibility $\chi_t$ (rhs.\ bottom) 
on different volumes $V=L^2$. We observe that the locations of the peaks of $\chi_c$ and $\chi_t$ approximately 
coincide and approach each other in the thermodynamic limit.  

In our upcoming study \cite{WIP2} we will present a detailed finite size scaling analysis of our observables across the 
three different branches of the phase boundaries in Fig.~\ref{fig:phase_diagram}. Based on this analysis we attempt a 
determination of the type of phase transitions and compare the emerging physical picture at $\theta = \pi$ with the
situation without topological term, i.e., the model at $\theta = 0$.

\section{Abelian topological terms in 4-d}

In this section we now have a brief look at the lattice
construction of the abelian topological charge 
via the construction \cite{Sulejmanpasic:2019ytl} in 4 dimensions \cite{WIP1}. 
The corresponding continuum expression is 
\begin{equation}
Q \; = \; \frac{1}{32\pi^2} \int \! d^4x \;
\varepsilon_{\mu \nu \rho \sigma} \, F_{\mu\nu}(x) \,
F_{\rho \sigma}(x) \; .
\label{Q4dcont}  	
\end{equation}
The strategy for the construction of the 4-d
topological charge will be as in the 2-d case, namely 
to use our lattice definition of the field strength
from Eq.~(\ref{Fmunudef}), i.e., 
$F_{x,\mu\nu} = (\mathrm{d}\, a)_{x,\mu \nu} + 
2\pi \, n_{x,\mu \nu}$ in a suitable discretization 
of the continuum form (\ref{Q4dcont}). 

However, before we can write down the lattice 
expression for $Q$ we need to have another look at 
the role of the Villain variables 
$n_{x,\mu \nu} \in \mathds{Z}$ in 4 dimensions. 
The $n_{x,\mu \nu}$ were 
introduced to remove the contributions 
$(\mathrm{d}\, k)_{x,\mu \nu}$ to the field strength 
$F_{x,\mu \nu}$, which emerge from
the shifts $2\pi k_{x,\mu}$ in the parameterization 
(\ref{Aparam}) of the non-compact fields $A_{x,\mu}$.
Since the $n_{x,\mu \nu}$ only need to remove 
terms of the form $(\mathrm{d}\, k)_{x,\mu \nu}$, the
$n_{x,\mu \nu}$ do not need to be general but can be 
restricted themselves. In particular we can implement 
the so-called {\sl closedness constraint} 
\begin{equation}
(\mathrm{d} \, n)_{x,\mu\nu\rho} \; = \; 0 
\quad \forall x, \; \mu < \nu < \rho \; 	,
\label{n_closed}
\end{equation}
since also the terms $(\mathrm{d}\,k)_{x,\mu \nu}$ obey
$(\mathrm{d} (\mathrm{d}\,k))_{x,\mu \nu \rho} = 0$ 
because of $\mathrm{d}^2 = 0$. 
Here we use the general definition of the exterior 
derivative d (see, e.g., \cite{Seiler,Wallace}) that 
turns a field $f_{x,\mu_1 \, ... \, \mu_{r-1}}$ 
with $r-1$ indices (a so-called $(r-1)$-form) 
into a field 
$(\mathrm{d}\, f)_{x,\mu_1 \, ... \, \mu_{r}}$ 
with $r$ indices (an $r$-form), 
\begin{equation}
(\mathrm{d}\,f)_{x,\,\mu_1 \mu_2 \, ... \, \mu_r} 
\; \equiv \;  
\sum_{j = 1}^r (-1)^{j+1} \, \Big[ \,f_{x+\hat{\mu}_j, \,\mu_1 \, ... \, {\mu}^{\!\!\!\!\!\!\!^o}_j  \, ... \, \mu_r} \, - \, 
f_{x, \,\mu_1 \, ... \, {\mu}^{\!\!\!\!\!\!\!^o}_j \, ... \, \mu_r} \,\Big] \; ,
\label{d_exterior}
\end{equation}
where ${\mu}^{\!\!\!\!\!\!\!^o}_j$ indicates that 
$\mu_j$ has been dropped from the list of indices.
It is easy to show the identity $\mathrm{d}^2 = 0$
from the definition (\ref{d_exterior}). 

The closedness constraint (\ref{n_closed})
requires that for all 3-cubes $(x,\mu < \nu < \rho)$ 
of the lattice the oriented sum over the Villain 
variables $n_{x,\mu \nu}$ on the plaquettes of the 
cube vanishes\footnote{Note that in 2-d
we do not have cubes, such that in Section 2 there is 
no equivalent of the discussion here in 4-d.}. 
We remark that the closedness constraint 
(\ref{n_closed}) corresponds to the absence
of monopoles, an aspect that we will discuss 
in more detail below.

Taking into account the closedness constraint we may formulate the partition sum of pure U(1) gauge theory in the form 
\begin{equation}
Z\; = \; 	\sum_{\{ n \}} \; \int \!\! D[a] \;\; \prod_{x} \prod_{\mu < \nu < \rho} \!  \delta\Big( (\mathrm{d}\,n)_{x,\mu \nu \rho}\Big) \;
e^{\, - \, \frac{\beta}{2} \sum_{x,\mu < \nu} \big( (\textrm{\scriptsize d} \, a)_{x,\mu \nu} \; + \; 2 \pi \, n_{x,\mu \nu}\big)^2} \; ,
\label{Z_lat_4d}
\end{equation}
where the measure $\sum_{\{ n \}} \; \int \! D[a]$ 
was defined in (\ref{measures}), and the closedness constraint has been implemented with a product of
Kronecker deltas over all cubes, where again we 
denote the Kronecker delta with $\delta(j) \equiv \delta_{j,0}$.

Having augmented the 4-d partition sum with the closedness constraint for the Villain variables we can 
now define the topological charge in our  
formulation \cite{Sulejmanpasic:2019ytl}, 
\begin{equation}
Q[F] \; = \; Q[\mathrm{d}\,a + 2\pi \, n]  \; \equiv \; \frac{1}{8 \pi^2} \sum_x \sum_{\mu < \nu \atop \rho < \sigma} 
F_{x,\mu \nu} \, \epsilon_{\mu \nu \rho \sigma} \, F_{x-\hat{\rho} - \hat{\sigma},\rho \sigma} \; ,
\;\; F_{x,\mu\nu} \; = \; (\mathrm{d}\, a)_{x,\mu \nu}
\, + \, 2\pi \, n_{x,\mu \nu} \; .
\label{Q4def}
\end{equation}
Obviously the definition of the topological charge 
is a simple discretization of the continuum 
expression (\ref{Q4dcont}) with the only subtle 
point that the second field strength vector is 
shifted by $-\hat\rho - \hat \sigma$, which
has the interpretation of a product 
of $F_{x,\mu \nu}$ on a plaquette 
$(x,\mu \nu)$ with the corresponding dual field 
strength on the plaquette dual to $(x,\mu \nu)$
\cite{Sulejmanpasic:2019ytl}. Note that we 
have partially ordered the sum of the Lorentz indices 
which alters the pre-factor. And we also stress once
more, that the definition of the topological charge 
(\ref{Q4def}) is understood with the condition that the 
Villain variables obey the closedness constraint. 

Using the closedness constraint we can now discuss features of $Q[F]$ and show that it has indeed the topological
properties we expect. In a straightforward but somewhat lengthy calculation \cite{WIP1} one can show that 
$Q[F]$ is invariant when adding an arbitrary 
exterior derivative $(\mathrm{d}\,B)_{x,\mu \nu}$ to the field strength $F_{x,\mu \nu}$,
\begin{equation}
Q[F + \mathrm{d} B] \; = \; Q[F] \; ,
\label{property1}
\end{equation}
where $B_{x,\mu}$ is an arbitrary link-based field. 
It is important to stress that this property only holds 
when the Villain variables $n_{x,\mu \nu}$ in 
$F_{x,\mu \nu} = (\mathrm{d}\, a)_{x,\mu \nu} + 2\pi \, n_{x,\mu \nu}$ obey the 
closedness condition (\ref{n_closed}).

The result (\ref{property1}) implies that 
$Q[F]$ is independent of the exterior derivative 
$(\mathrm{d}\,a)_{x,\mu \nu}$ in $F_{x,\mu \nu} = (\mathrm{d} \, a)_{x,\mu \nu} + 2\pi \, n_{x,\mu \nu}$, which can be seen by 
setting $B_{x,\mu} = -a_{x,\mu}$. As a consequence, 
like in the 2-d case, also in 4-d the topological charge depends 
only on the Villain variables $n_{x,\mu \nu}$ 
and we find (under the assumption 
$(\mathrm{d}\,n)_ {x,\mu \nu \rho} \; = \; 0$),
\begin{equation}
Q[F] \; = \; Q[2 \pi \, n]  \; = \; \frac{1}{2} \sum_x \sum_{\mu < \nu \atop \rho < \sigma} 
n_{x,\mu \nu} \, \epsilon_{\mu \nu \rho \sigma} \, n_{x-\hat{\rho} - \hat{\sigma},\rho \sigma} \; .
\label{Qnn}
\end{equation}  
To establish that $Q[2 \pi \, n]$ is integer-valued we now 
invoke the Hodge decomposition 
(see, e.g., \cite{Seiler,Wallace}) which states that 
an arbitrary $r$-form can be written as the sum 
of the exterior derivative of an $(r-1)$-form, the 
boundary of an $(r+1)$-form and a harmonic $r$-form 
$h$ which obeys $\mathrm{d}\, h = 0$. Using the fact that
the Villain variables obey 
$(\mathrm{d}\,n)_{x,\mu \nu \rho} = 0$ and thus have no 
boundary contribution, we can write our 
Villain variables as
\begin{equation} 
n_{x,\mu \nu} \; = \; (\mathrm{d} \, l)_{x,\mu \nu} 
\; + \; h_{x,\mu \nu} \; ,
\label{hodge}
\end{equation}
where $l_{x,\mu} \in \mathds{Z}$ is an integer-valued link field, which 
due to $\mathrm{d}^2 \, l = 0$ does not contribute to $\mathrm{d} \, n$. 
The harmonic contributions can be written in the form ($\mu < \nu$)
\begin{equation}
h_{x,\mu \nu} \; = \; \omega_{\mu \nu} \, \sum_{i=1}^{N_\rho} \sum_{j=1}^{N_\sigma} 
\delta_{x,i \hat\rho + j \hat\sigma}^{\; (4)} 
\; \; \; \; \; \mbox{with} \; \; \; \; \; \rho \neq \mu,\nu \, ; \; \;  \sigma \neq \mu,\nu \, ; \; \;  \rho \neq \sigma \; .
\label{harmonics}
\end{equation}
$N_\rho$ and $N_\sigma$ are the lattice extents in the $\rho$- and $\sigma$-directions and 
$\delta_{x,y}^{\; (4)}$ denotes the 4-dimensional Kronecker delta. Thus
$h_{x,\mu \nu}$ vanishes everywhere except on all $\mu$-$\nu$ plaquettes  
$(x,\mu \nu)$ that have their root site $x$ in the $\rho$-$\sigma$ plane that is orthogonal to the $\mu$-$\nu$ plane 
and contains the origin, where it has a constant value $\omega_{\mu \nu} \in \mathds{Z}$. 
For all other plaquettes $h_{x,\mu \nu} = 0$. It is easy to check that the harmonics 
are closed forms, i.e., $(\mathrm{d}\,h)_{x,\mu\nu\rho} = 0$, and that they can not be written 
as an exterior derivative of an $(r-1)$-form. The choice of the $h_{x,\mu \nu}$ 
is not unique, as they can always be deformed by adding an arbitrary exterior derivative. 

Inserting the representation (\ref{hodge}), (\ref{harmonics}) into the topological charge definition (\ref{Qnn})  one finds
\begin{equation}
Q[2\pi \, n] \; = \; Q[2 \pi \, h] \; = \; \omega_{12} \, \omega_{34} \; - \; \omega_{13} \, \omega_{24} \; + \;
\omega_{14} \, \omega_{23} \; = \; \frac{1}{8} \epsilon_{\mu \nu \rho \sigma} \, \omega_{\mu \nu} \, \omega_{\rho \sigma} 
\; \in \; \mathds{Z} \; ,
\label{Qharmonics}
\end{equation}  
where the first part is a direct consequence of (\ref{property1}) and implies that the topological charge depends 
only on the harmonic contributions. Inserting the harmonic contribution (\ref{harmonics}) into $Q$ 
one then may work out the explicit form on the rhs.\ of (\ref{Qharmonics}) which also establishes that
$Q$ is integer. The result (\ref{Qharmonics}) completes our analysis of the topological charge, 
which for closed Villain variables obeys $Q[F] = Q[2 \pi \, n] = Q[2 \pi \, h]$ and shows that the topological charge is 
integer-valued and is uniquely determined by the harmonics in the Hodge decomposition of the Villain variables. 

Using our construction of the topological charge we can define the partition sum for U(1) lattice gauge theory 
with a topological term, 
\begin{equation}
Z  =  \sum_{\{ n \}} \int \! \! D[a] \,  
\exp\!\left( \!\!-\frac{\beta}{2} \sum_{x,\mu < \nu} \! \! F_{x,\mu \nu} F_{x,\mu \nu} \, + \, i  
\frac{\theta}{2} \sum_x \sum_{\mu < \nu \atop \rho < \sigma} 
n_{x,\mu \nu} \, \epsilon_{\mu \nu \rho \sigma} \, n_{x-\hat{\rho} - \hat{\sigma},\rho \sigma} \! \!\right) \!
\prod_{x} \! \prod_{\mu < \nu < \rho}  \!\!\!\!\! \delta\Big( \! (\mathrm{d}\,n)_{x,\mu \nu \rho}\!\Big),  
\label{Zdef2}
\end{equation}
with $F_{x,\mu \nu} \; = \; (\mathrm{d}\,A + 2\pi n)_{x,\mu\nu}$. 

It is interesting to observe the structural similarity but also the differences between the 4-d partition sum 
(\ref{Zdef2}) and its 2-d counterpart (\ref{Z_lat_2d}). In both cases the gauge field action is the usual 
$F_{\mu \nu}^{\; 2}$ term and the topological charge depends only on the Villain variables. However, in 4-d this 
requires the implementation of the closedness constraint which removes monopoles, which in 2-d has no analogue.

For further  understanding the roles of the monopoles we now discuss an interesting physical 
aspect of the abelian theta term which is also an important consistency check of our formulation, the so-called 
Witten effect. The Witten effect states that the theta term endows a magnetic monopole with the minimal 
possible magnetic charge 
$m = 1$ with an electric charge $q = \theta/2\pi$. 

In order to place a single monopole of magnetic charge $m = 1$ we relax the closedness constraint on a single 
3-cube, and set\footnote{For an alternative demonstration that the topological charge based on our construction 
correctly implements the  Witten effect see \cite{Sulejmanpasic:2019ytl}.}
\begin{equation}
(\mathrm{d} \, n)_{x,\, \nu \rho \sigma} \; = \; \delta^{(4)}_{x,\, x_0 + \hat{2} + \hat{3} + \hat{4} } \; 
\delta_{\nu,2} \, \delta_{\rho,3} \, \delta_{\sigma,4} \; . 
\label{dn_monopole}
\end{equation}
The exterior derivative vanishes everywhere, except for the single 3-cube with orientation $\nu = 2$, $\rho = 3$, 
$\sigma = 4$ and root site $x_0 + \hat{2} + \hat{3} + \hat{4}$, where 
the shift $\hat{2} + \hat{3} + \hat{4}$ was chosen for notational convenience. 
Evaluating the topological charge with this choice of constraints for the Villain variables one finds 
(here we use the abbreviation $\hat{s} \equiv \hat1 + \hat2 + \hat3 + \hat4$)
\begin{eqnarray}
Q[F] & \! = \! & \frac{1}{8 \pi^2} \sum_x \sum_{\mu < \nu \atop \rho < \sigma} 
(\mathrm{d} \, A + 2\pi n)_{x,\mu \nu} \, \epsilon_{\mu \nu \rho \sigma} \, (\mathrm{d}\,A + 2\pi n)_{x-\hat{\rho} - \hat{\sigma},\rho \sigma} 
\\
& \! = \! & Q[\mathrm{d}\, A] \; + \; Q[2 \pi \, n] \; + \; \frac{1}{4 \pi} \sum_x \sum_{\mu < \nu \atop \rho < \sigma}
\Big[(\mathrm{d}\,A)_{x,\mu \nu} \, \epsilon_{\mu \nu \rho \sigma} \, n_{x-\hat{\rho} - \hat{\sigma},\rho \sigma} \, + \,
n_{x,\mu \nu} \, \epsilon_{\mu \nu \rho \sigma} \, (\mathrm{d}\, A)_{x-\hat{\rho} - \hat{\sigma},\rho \sigma}\Big]
\nonumber \\
& \! = \! & Q[2 \pi \, n] \; + \; \sum_{x}  \sum_{\mu < \nu \atop \rho < \sigma}
\frac{A_{x+\hat{s} - \hat{\mu},\,\mu} +  A_{x- \hat{\mu},\,\mu}}{4 \pi} \, \epsilon_{\mu \nu \rho \sigma} \, 
(\mathrm{d}\,n)_{x,\, \nu \rho \sigma} \; = \; Q[2 \pi \, n] \; + \; \frac{1}{2\pi} \frac{A_{x_0,\,1} +  A_{x_0-\hat{s},\,1}}{2} \; ,
\nonumber
\end{eqnarray}
where in the first step we have isolated the mixed terms from the quadratic ones and in the second step used
$Q[\mathrm{d}\,A] = 0$. In the third step we used a straightforward but lengthy calculation \cite{WIP1}, and in the last step 
we inserted the choice (\ref{dn_monopole}) for the constraints. Thus when adding the topological term 
$i \, \theta \, Q[F]$ to the action, the monopole generates the term 
\begin{equation}
i \, \frac{\theta}{2 \pi} \, \frac{A_{x_0,\,1} +  A_{x_0-\hat{s},\,1}}{2} \; \equiv \; i \, q \; \bar{A}_{x_0,\,1} 
\qquad \mbox{with} \quad q \; = \; \frac{\theta}{2 \pi} \; , \; \; 
\bar{A}_{x_0,\,1} \; = \; \frac{A_{x_0,\,1} +  A_{x_0-\hat{s},\,1}}{2} \; .
\end{equation}
Obviously the magnetic monopole couples to the gauge field, which here appears as the averaged 
1-component $\bar{A}_{x_0,\,1}  = (A_{x_0,\,1} +  A_{x_0-\hat{s},\,1})/2$, with a charge $q = \theta/2\pi$,
as expected for the Witten effect. This result constitutes an important consistency check for our 
approach to the lattice discretization of the topological charge. 

Having identified a suitable lattice discretization of the 4-d abelian topological charge, the whole formalism will be
developed further in our upcoming paper \cite{WIP1}. We will show that the topological term can be 
generalized further and extended to a whole family of different discretizations of the topological charge. Based 
on this generalization it is possible to construct a self-dual 4-d U(1) lattice gauge theory with a topological term
that is obtained as a linear superposition of the members of the family of topological charges. 
This clearly is a highly interesting finding, since self-duality can be used to obtain non-perturbative 
insight into observables in the presence of the topological term. In yet another generalization step we couple 
bosonic matter fields to the gauge fields and find that if electrically and magnetically charged field are coupled 
in a symmetrical way, also the gauge-Higgs model with a topological term becomes fully self-dual.

\section{Summary and outlook}

In this proceedings contribution we have presented an overview of results from our program 
\cite{Gattringer:2018dlw,Goschl:2018uma,Sulejmanpasic:2019ytl,Gattringer:2019yof,WIP1,WIP2} where we consider 
the lattice discretization of topological terms in abelian theories. Our approach is based on a  
formulation \cite{Sulejmanpasic:2019ytl} of the U(1) lattice gauge theory which leads to a Villain action that in the presence
of a topological term is generalized such that also the topological charge $Q$ is taken into account. We show that 
for both, 2-d and 4-d the topological charge is a function of only the integer-valued Villain variables, where in the 4-d
case a closedness constraint for the Villain variables was used to remove the abelian monopole contributions. 

Part of this overview presentation are tests of the 2-d and 4-d topological charge definitions. For the 2-d case we show 
that two different definitions of the topological charge based on the formulation in \cite{Sulejmanpasic:2019ytl} both obey the 
index theorem in the continuum limit. This is an important check for a lattice discretization of the topological charge
since in a gauge theory with fermions the index theorem is the link 
between topological excitations and fermion phenomenology. 
For the 4-d case we show that the topological charge receives contributions only from the harmonic part of the 
Hodge decomposition of the Villain variables, which establishes the topological nature of our discretization of $Q$.
Furthermore we show that the Witten effect is implemented correctly, implying that in the presence of a topological term
a monopole with magnetic charge $m = 1$  receives an electric charge of $q = \theta/2\pi$. 

Besides the construction of the topological charge and studies of its properties we also present results of an 
application of our new formulation for the topological charge 
in a Monte Carlo analysis of the 2-flavor U(1) gauge Higgs model at topological angle $\theta = \pi$.
The simulation is based on a worldline/worldsheet representation which overcomes the complex action problem. 
We present first results for the phase diagram of the model whose structure is determined by the breaking of the 
non-trivial charge conjugation symmetry at $\theta = \pi$, as well as breaking of the flavor symmetry when varying 
a relevant deformation parameter in the 2-flavor self-interaction. A more detailed analysis of the types of 
transitions across the phase boundaries is in preparation \cite{WIP2}. 

There are several possible directions for further studies, which we partly have already started to work on. As already 
outlined in the end of Section 6, in 4 dimensions one may use a more general definition of the topological 
charge to construct 
a fully self-dual U(1) lattice gauge theory with a topological term \cite{WIP1}. This theory can be extended to a gauge 
Higgs model with electric as well as magnetic matter which again can be shown to be fully self-dual 
\cite{Sulejmanpasic:2019ytl,WIP1}, and we have started to explore the consequences of self-duality also with 
numerical simulations. Yet another interesting case to study are U(1) gauge Higgs models in 3 dimensions, where it was
shown \cite{Sulejmanpasic:2019ytl} that with our discretization we may control the charge of monopoles 
and study the corresponding physics with Monte Carlo simulations. Also for this application of our formalism numerical
studies are planned for the near future. 

\vskip5mm
\noindent
{\bf Acknowledgments:}
\vskip2mm
\noindent 
TS is supported by the Royal Society University Research Fellowship. CG acknowledges support by the 
Dr.\ Heinrich J\"org foundation, Karl-Franzens-University Graz.
Furthermore this work is supported by the Austrian Science Fund FWF, grant I 2886-N27 and the FWF
DK ''Hadrons in Vacuum, Nuclei and Stars'', grant W-1203. 
Parts of the computational results presented here have been achieved using the Vienna Scientific Cluster (VSC).


\begin{thebibliography}{99}

\bibitem{Gattringer:2018dlw}
  C.~Gattringer, D.~G{\"o}schl, T.~Sulejmanpasic,
  Nucl.\ Phys.\  {\bf B 935} (2018) 344
  [arXiv:1807.07793].
  
\bibitem{Goschl:2018uma}
  D.~G{\"o}schl, C.~Gattringer, T.~Sulejmanpasic,
  PoS LATTICE2018 (2019) 226 [arXiv:1810.09671].

\bibitem{Sulejmanpasic:2019ytl}
  T.~Sulejmanpasic, C.~Gattringer,
  Nucl. Phys. {\bf B 943C} (2019) 114616 [arXiv:1901.02637].
  
\bibitem{Gattringer:2019yof}
  C.~Gattringer, P.~T{\"o}rek,
  Phys.\ Lett.\ B {\bf 795} (2019) 581
[arXiv:1905.03963].

\bibitem{WIP1}
  M.~Anosova, C.~Gattringer, T.~Sulejmanpasic, work in preparation. 
  
\bibitem{WIP2}
 C.~Gattringer, D.~G{\"o}schl, T.~Sulejmanpasic, work in preparation. 
  
\bibitem{Gaiotto:2017yup}
  D.~Gaiotto, A.~Kapustin, Z.~Komargodski, N.~Seiberg,
  JHEP {\bf 1705} (2017) 091
  [arXiv:1703.00501].
  
\bibitem{Villain:1974ir}
  J.~Villain,
  J.\ Phys.\ (France) {\bf 36} (1975) 581.
  
\bibitem{Atiyah}
   M.~Atiyah, I.M.~Singer, 
   Ann.\ Math.\ {\bf 93} (1971) 139.
      
\bibitem{vanishing1}
J.\ Kiskis, Phys. Rev. {\bf D 15} (1977) 2329.

\bibitem{vanishing2}
N.K.\ Nielsen, B.\ Schroer, Nucl.\ Phys.\ {\bf B 127} (1977) 493.

\bibitem{vanishing3} 
M.M.\ Ansourian, Phys.\ Lett.\ {\bf 70 B} (1977) 301.
    
\bibitem{overlap1}
H.~Neuberger, Phys.\ Lett.\ {\bf B 417} (1998) 141.

\bibitem{overlap2}
H.~Neuberger, Phys.\ Lett.\ {\bf B 427} (1998) 353.

\bibitem{indexlattice}
M.\ L{\"u}scher, P.\ Weisz, JHEP {\bf 0207} (2002) 049.

\bibitem{Sulejmanpasic:2018upi} 
  T.~Sulejmanpasic and Y.~Tanizaki,
  Phys.\ Rev.\ B {\bf 97}, no. 14, 144201 (2018)


\bibitem{Tanizaki:2018xto} 
  Y.~Tanizaki and T.~Sulejmanpasic,
  Phys.\ Rev.\ B {\bf 98}, no. 11, 115126 (2018)

    
\bibitem{Kloiber:2014dfa}
  T.~Kloiber, C.~Gattringer,
  PoS LATTICE {\bf 2014} (2014) 345
  [arXiv:1410.3216]. 

\bibitem{Gattringer:2015baa}
  C.~Gattringer, T.~Kloiber, M.~M{\"u}ller-Preussker,
  Phys.\ Rev.\ D {\bf 92} (2015) 114508
  [arXiv:1508.00681].
    
\bibitem{Mercado:2013yta}
  Y.~Delgado Mercado, C.~Gattringer, A.~Schmidt,
  Comput.\ Phys.\ Commun.\  {\bf 184} (2013) 1535
  [arXiv:1211.3436].

\bibitem{Schmidt:2012uy}
  A.~Schmidt, Y.~Delgado Mercado, C.~Gattringer,
  PoS LATTICE {\bf 2012} (2012) 098
  [arXiv:1211.1573].

\bibitem{Mercado:2013ola}
  Y.~Delgado Mercado, C.~Gattringer, A.~Schmidt,
  Phys.\ Rev.\ Lett.\  {\bf 111} (2013) 141601
  [arXiv:1307.6120].
  
\bibitem{Mercado:2013jma}
  Y.~Delgado, C.~Gattringer, A.~Schmidt,
  PoS LATTICE {\bf 2013} (2014) 147
  [arXiv:1311.1966].

\bibitem{Seiler}
E.~Seiler,
  {\sl Gauge Theories as a Problem of Constructive Quantum Field Theory and Statistical Mechanics},
  Lect.\ Notes Phys.\  {\bf 159} (1982) 1.

\bibitem{Wallace}
A.H.~Wallace, {\sl Algebraic Topology, Homology and Cohomology}, W.A. Benjamin, New York 1970.

\end{thebibliography}
\end{document}